\documentstyle[12pt]{article}
\begin{document}
\title{Generation of Source Terms in General Relativity by
  differential structures}
\author{T. Asselmeyer\thanks{torsten@summa.physik.hu-berlin.de} \\ 
Institut of Physics, Humboldt University \\ Berlin, Germany}
\date{October 9, 1996}
\maketitle
\begin{abstract}
  In this paper the relation between the choice of a differential
  structure and a smooth connection in the tangential bundle is
  discussed. For the case of an exotic $S^7$ one obtains corrections
  to the curvature after the change of the differential structure,
  which can not be neglected by a gauge transformation. 
  In the more interesting case of four dimensions we obtain a
  correction of the connection constructed by intersections of
  embedded surfaces. This correction produce a source term in the
  equation of the general relativity theory which can be interpreted
  as the energy-momentum tensor of a embedded surface.
\end{abstract}
{\bf Pacs: 04.20.Gz, 02.40.Re, 02.40Vh}\\[0.3cm]
{\it submitted to Classical and Quantum Gravity}

\section{Introduction}
One of the outstanding problems in physics is the unification of the
quantum and general relativity theory. As we learned from gauge field
theory and from the mathematical part known as geometric quantization,
the procedure of quantization leads to the introduction of topological
invariants defined on the phase space of the systems.  In the case of
the quantization of gravity one works on the space of
(pseudo-)Riemannian metrics or better on the space of connections
(Levi-Civita or more general). The diffeomorphism group of the
corresponding manifold is the gauge group of this theory.  Knowing
more about the structure of this group means knowing more about
quantum gravity.  In a first approximation we are interested in the
mathematical question: If one smooth manifold is homeomorphic to
another one, need it be diffeomorphic? In 1956 J. Milnor \cite{Mil:56}
gave a negative answer to this question. He constructed the first
exotic manifold, the exotic $S^7$, as the total space of a fiber
bundle. After this remarkable result, the next important step was the
proof of the generalized Poincar\'e conjecture for dimensions greater
than 4, also known as h-cobordism theorem \cite{Mil:65}.  Together
with the calculation of the h-cobordism group $\Theta_n$ of homotopy
$n$-spheres by Kervaire and Milnor \cite{KerMil:63} one obtains the
classification of differential structures on topological $n$-spheres.
Later Kirby and Siebenman \cite{KirSie:77} proved that every simply
connected manifold of dimension greater than 4 can be classified by
$\Theta_n$. By obstruction theory one obtains an element of the
cohomology group $H^{n+1}(M,\Theta_n)$ of the manifold $M$ which
measures the existence of the differential structures. The rank of
$H^n(M,\Theta_n)$ gives the number of such structures. Because of the
failure of Whitneys trick, only the special case of dimension 4
remained unsolved for a long time. With the help of the technique of
Casson handles, Freedman \cite{Fre:82} proved in 1982, that the
classification of topological 4-manifolds is given by the
classification of the so-called intersection form. Soon afterwards
Donaldson \cite{Don:83} published his famous theorem, which shows the
existence of an exotic ${I\mkern-6.0mu R}^4$, which is impossible for
the ${I\mkern-6.0mu R}^n$ with $n\not=4$. In contrast to the compact
case there are uncountable many exotic ${I\mkern-6.0mu R}^4$
\cite{Gom:85,Tau:87}.  So the most interesting dimension for physics
has the richest structure. Now the question is: What is the physical
relevance of the differential structure? Which physical observable
will be modified after a change of the differential structure? In the
case of the exotic ${I\mkern-6.0mu R}^4$ the first question was
discussed by Brans and Randall \cite{BraRan:93} and later by Brans
\cite{Bra:94a,Bra:94b} alone to guess, that exotic smoothness can be a
source of non-standard solutions of Einsteins equation. In connection
with the topological quantum field theory other physical models are
developed to relate the topological invariants to vacuum states of the
quantum field theory.  In this paper we try to give an answer to the
second question. To this end, we introduce a toy model given by the
exotic $S^7$ and study the change of the connection in the tangent
bundle with respect to the change of the differential structure.  The
fact behind this construction is the localization of exotic smoothness
which means that the ``exoticness'' of the structure is induced by a
countable set of points on the manifold. In the compact case this can
be described by certain characteristic classes as was done in the
first paper of Milnor \cite{Mil:56}, or by generalized characteristic
classes given by the bigraded version of the Weil-algebra
\cite{Dub:94}. In the non-compact case, as in ${I\mkern-6.0mu R}^4$,
such powerful algebraic invariants are missing. Otherwise the compact
case has physically more sense and we will investigate this case in
detail.

In the next section we construct explicitly an exotic $S^7$ following
Milnor \cite{Mil:56} and choose a connection in the tangent bundle.
Then we define a map between the standard and the exotic $S^7$ which
is a differentiable map except for one point. The associated map
between the tangent bundles possesses a singularity originating from
the fact, that the exotic $S^7$ has non-trivial first Pontrjagin
class. From the theory of Harvey and Lawson \cite{HarLaw:93} we know
that this singularity is related to the occurrence of a nontrivial
Pontrjagin class and is explicitly realized as current. So we obtain
the nice result that a change of the differential structure induces a
change in the connection which can not be neglected by a
diffeomorphism. The four-dimensional case is related to the Donaldson
polynomial \cite{Don:90,DonKro:90} as topological invariant to
classify the differential structures for the compact 4-manifolds $M$.
Because of the realization as polynomial in $H_2(M)$, an associated
change in the connection is discussed. The new invariants of Seiberg
and Witten \cite{SeWi:94,Wit:94} use also the relation between the
elements in $H_2(M)$. So we construct with the help of these classes
in $H_2(M)$ the change of the connection after the change of the
differential structure. To this end we use the results of Kronheimer
and Mrowka \cite{KrMr:94} for manifolds of simple type that there is
$d$ classes $e_1,\ldots,e_d\in H^2(M,{Z\!\!\!Z})$ and $d$ polynomials
$f_1,\ldots, f_d\in\mbox{ \leavevmode\hbox{\kern.3em\vrule
     height 1.2ex depth -.3ex width .2pt\kern-.3em\mbox{\rm Q}}}[z]$ such that the Donaldsons invariant is
uniquely defined by these pairs. Furthermore the set of these classes
$\{ e_i\}$ with $i=1,\ldots,k$ is a diffeomorphism invariant. So one
can conclude that two different differential structures differ by the
number of classes in $H^2(M,{Z\!\!\!Z})$. From this knowledge we construct the
change of the connection by changing the differential structure. A
physical interpretation of this mathematical construction is given by
the general relativity theory as the energy-momentum tensor of a
embedded surface in the 4-manifold which can be interpreted as a world
surface of a string after the choice of one coordinate as time
coordinate. 

\section{The simplest example - the exotic $S^7$}
\label{sec:2}
The change of the differential structure will be understanded in 
the following sense. We choose a topological, smoothable manifold
$M$ with the micro tangential bundle $\tau(M)$ \cite{Mil:64}. From
this bundle we get back two different tangential bundles $TM$ and
$TM^\prime$ associated to different differential structures. The
change of differential structure means that the following diagram of
homeomorphisms commutes
\begin{eqnarray*}
  \begin{array}{ccc}
    TM & & \\ & \searrow & \\ d\alpha\;\downarrow & & \tau(M) \\ & \swarrow &
    \\ TM^\prime & &
  \end{array}\quad .
\end{eqnarray*}
The map $\alpha$ is a diffeomorphism except on a countable set of
points. In the case of $M=S^7$ we know that a smooth structure exists
which is not unique. To construct and detect such structure we refer
to \cite{Mil:56}. 

Let $S^7$ be the standard $S^7$ and $M^7_k$ one of the exotic $S^7$.
Furthermore we assume that a Riemannian
metric is chosen on both manifolds. Let $D_{TS^7}$ be a connection in
the tangential bundle $TS^7$ and $D_{TM^7_k}$ in $TM^7_k$. 
We are interested in the construction of the difference connection
$D_{dc}$ given by  
\begin{equation}
D_{TM^7_k}=D_{TS^7}+D_{dc} \quad \mbox{in $S^7$} \qquad.
\end{equation} 
The reason for the introduction is the effect, that the transport of a
section in $M^7_k$ is different from the transport of the same section
in $S^7$ because the section smooth in $M^7_k$ is non-smooth in some
points in $S^7$. The difference connection $D_{dc}$ compensate this
effect. To construct this connection we investigate the map
$h:M^7_k\longrightarrow S^7$. This map is up to one point 
$x_0\in M^7_k$ a diffeomorphism. So the map $dh:TM^7_k\longrightarrow TS^7$ is
singular with respect to this point $x_0$. Next we fix a frame $e$ in
$TM^7_k$ and in $TS^7$ denoted by $f$. Together with the map we obtain
$dh(e)=a\cdot f$ where $a$ is a $SO(7)$-valued function\footnote{The
  function has in general values in $Gl(7,{I\mkern-6.0mu R})$, but this group can be
  reduced to the $SO(7)$ group by bundle reduction. Furthermore we
  assume without loss of generality that the space has a euclidian
  metric.} with a singularity on the same place. By the standard
procedure of gauging the connection, the difference connection is
given by 
\begin{equation} 
D_{dc}=a^{-1}da 
\end{equation} 
with the property 
\begin{equation}
d(a^{-1}da)+a^{-1}da\wedge a^{-1}da=0 \quad \mbox{in}\quad
T(M^7_k\backslash \{x_0\}) \qquad .
\end{equation} 
Loosely speaking the curvature
induced by changing the differential structure is concentrated on the
point $x_0$ and one can write 
\begin{equation} 
d(a^{-1}da)+a^{-1}da\wedge a^{-1}da=a(x_0)\cdot\delta (x_0)
\label{sing}\quad \mbox{in}\quad TM^7_k 
\end{equation} 
where $\delta(x_0)$ is the delta distribution and $a(x_0)\in
SO(7)$ is the group element corresponding to the point $x_0$. To
clarify this important point in the argumentation, we choose without
loss of generality a coordinate system with $x_0=0$. Together with the
parallelization of the topological $S^7$ we can split the differential
$dh:TM^7_k\longrightarrow TS^7$ to get $dh=(h,a):M^7_k\times{I\mkern-6.0mu R}^7\longrightarrow
S^7\times{I\mkern-6.0mu R}^7$ with the same map $a$ defined above. The group element
$a\in SO(7)$ is considered with respect to the map $h$ which means
that the $a(0)$ corresponding to the point $0$ is singular. Milnor 
\cite{Mil:56} constructed the homeomorphism $h$ with a Morse function 
$f:M^7_k\longrightarrow{I\mkern-6.0mu R}$ having two non-degenerated, critical points $x_0,x_1$. 
Next we calculate the tangential vector field corresponding to the Morse 
function:
\begin{equation} 
\frac{d\vec{x}(t)}{dt}=\nabla_{\dot{x}} f \label{dynamik}\qquad .
\end{equation}
According to Milnor we normalize the function to be $f(x_1)=0$ and 
$f(x_0)=1$. The solution of (\ref{dynamik}) induces the homeomorphism
$h$ which is a diffeomorphism except at one point $x_0$. Of course
at the critical points the tangential vector field vanish and the group
element $a(x_0)$ is defined to be the transformation $a(x_0)y=0$ where
$y\in{I\mkern-6.0mu R}^7$ is an element of the tangential space over $x_0$. The group 
element $a$ depends of the point in $M^7_k$ and can be considered as a
element of the gauge group ${\cal G}(M^7_k)$. This group is described via
the set of sections of the automorphism bundle $Aut(TM^7_k)$ and act on
every connection one-form $\omega\in T^*(TM^7_k)\otimes\underline{so}(7)$ on
$TM^7_k$, where $\underline{so}(7)$ is the Lie algebra of $SO(7)$.
The bundle $Aut(TM^7_k)$ is a bundle of $SO(7)$ groups and we construct
form this bundle a associated real vector bundle $V(TM^7_k)$ of rank 7.
The difference connection $D_{dc}$ is a homogeneous form over $V(TM^7_k)$ 
which means the invariance of the form after left multiplication with
a scalar. We denote the homogeneous one-form over $V(TM^7_k)$ associated
to the difference connection by $\omega\in T^*M^7_k\otimes T^*{I\mkern-6.0mu R}^7$ and
the coordinate system on every fiber by $y\in{I\mkern-6.0mu R}^7$. Next we use a theorem in
\cite{HarLaw:93} (chapter III) to get the formula for the exterior 
derivation of $\omega$ 
\begin{equation}
(d\omega)=(d\omega)_{regular}+\left(\int\limits_{S^1}\omega
\right)\frac{1}{\gamma_7}(d*d(\frac{1}{|y|^5})) \qquad\mbox{in}\quad V(TM^7_k) 
\end{equation} 
where $\gamma_7$
is the volume of the $S^6$ unit sphere. From the standard result
\[*d*d(\frac{1}{|y|^5})=\Delta_y\frac{1}{|y|^5}=\gamma_7\delta(y) \]
in functional analysis we get the outstanding delta function. Now we
pull-back this formula to $TM^7_k$ via the element $a$ of the gauge
group with $a(x_0)y=0$ to get
\begin{equation}
(da^*\omega)=(da^*\omega)_{regular}+\left(\,\int\limits_{S^1}a^*\omega\,
\right)\frac{1}{\gamma_7}(d*d(a^*(\frac{1}{|y|^5})))\quad\mbox{in}\quad
TM^7_k \label{change-S7}
\end{equation} 
Because of the property $a(x_0)y=0$
the support of the form $(d*d(a^*(\frac{1}{|y|^5})))$ is the single
point $x_0$ leading to the final result (\ref{sing}).  

According to \cite{KirSie:77} (Essay IV, Theorem 10.1) one can
consider the smoothing problem as lifting problem described by
obstruction theory \cite{Spa:66,GriMor:81}. In our case $M=S^7$ the
obstruction to smooth the $S^7$ is an element of
$H^8(S^7,{Z\!\!\!Z}_{28})$ (see \cite{KerMil:63}) which vanish in this
case. The number of smoothings are the number of elements in
$H^7(M^7_k,{Z\!\!\!Z}_{28})={Z\!\!\!Z}_{28}$. In equation
(\ref{change-S7}) the one-form $a^*\omega$ represents a non-trivial
cohomology class. The 7-form $(a^*\omega)^7$ can be interpreted as the
generator of the group $H^7(M^7_k,{Z\!\!\!Z})$. Using the relation 
\begin{equation}
\int\limits_{S^7} \delta(x_0)\, *1 = 1 
\end{equation} 
where $*$ is the Hodge operator, we obtain for the curvature 
\begin{equation} \Omega_{dc}=\frac{1}{28}\,\delta(x_0) a(x_0)\sum\limits_{i<j}
\frac{1}{21}dx^i\wedge dx^j 
\end{equation} 
with the normalization factor $1/28$.
This relates the obstruction to the change in the curvature after a
change in the differential structure. 

In the interesting case of a 4-manifold we have to use a more
sophisticated technique developed by Casson, Freedman and Donaldson to
get the same results. This will be done in the following section.

\section{Discussion of the four-dimensional case}
\subsection{Introducing Remarks}
\label{sec:3.1}
The four-dimensional case requires completely different methods than
the higher-dimensional one. At first we consider a compact 4-manifold
$M$. Now one may ask: What is the characterization of the differential
structures in this manifold? Fortunately the invariants of Donaldson
and Seiberg-Witten answer this question particularly. The main
components of the 4-manifold are formed by the basis elements of the
second homology group $H_2(M,{Z\!\!\!Z})$ except the generators of the torsion
subgroup. If we choose a simple-connected 4-manifold then the group
admits no torsion.  Now one can define a pairing
\begin{eqnarray*}
\begin{array}{cccc} \omega:& H_2(M,{Z\!\!\!Z})\times H_2(M,{Z\!\!\!Z}) & \longrightarrow & {Z\!\!\!Z} \\
   & (\alpha,\beta) & \longmapsto & <\alpha,\beta>\end{array}  
\end{eqnarray*}
where $<\alpha,\beta>$ denotes the intersection number of the two
cycles. This pairing is known as the intersection form and following
Freedman it is the main topological invariant to distinguish two
simple-connected, topological 4-manifolds. With the help of the
Poincar\'e dual we can switch to the picture of cohomology
classes. Fortunately there is a unique characterization of the classes
in $H^2(M,{Z\!\!\!Z})$ by complex line bundles over $M$. The position of the
cycles to each other, expressed by the topological properties of the
moduli space of an appropriated (non-linear) gauge theory, creates
diffeomorphism invariants to distinguish different differential
structures. Because of the failure of the h-cobordism theorem in four
dimensions we have to change the description from the Morse function
picture to methods which use the complex line bundle.

Let $P$ be a $SU(2)$-principal bundle over $M$ with second Chern
number $k$ and the moduli space of irreducible, anti-self-dual,
gauge-equivalent connections is denoted by ${\cal M}$. Let $\cal G$ be
the gauge group of $P$ and $\cal A$ the space of irreducible
connections. From the bundle $P$ one can induce the tautological
bundle $\underline{P}={\cal A}\times P$.  But in the case of
irreducible connections the gauge group $\cal G$ does not freely act
on $\cal A$ and one has to factorize out the center of the $SU(2)$
which is given by the group ${Z\!\!\!Z}_2$. This procedure results in a
universal $SO(3)$ bundle ${\cal P}^{ad}=\underline{P}/{\cal G}$ over
${\cal M}\times M$ with the first Pontrjagin class $p_1({\cal
  P}^{ad})$. Next we define the map $\mu:H_2(M,\mbox{ \leavevmode\hbox{\kern.3em\vrule
     height 1.2ex depth -.3ex width .2pt\kern-.3em\mbox{\rm Q}}})\longrightarrow H^2({\cal
  M},\mbox{ \leavevmode\hbox{\kern.3em\vrule
     height 1.2ex depth -.3ex width .2pt\kern-.3em\mbox{\rm Q}}})$ by
\[ \int\limits_{T} \mu(\Sigma)=-\frac{1}{4}\int\limits_{T\times[\Sigma]} 
p_1({\cal P}^{ad}) \] where $\Sigma$ is a 2-dimensional submanifold of
$M$ and $T\subset {\cal M}$ is a compact 2-dimensional submanifold of
the moduli space. Now we consider the situation, in which the moduli
space is even-dimensional $\dim {\cal M}_k=2d(k)$. This generalization
superimposes a restriction on the manifold $M$ but it does include all
interesting cases. We now choose $d$ classes $[\Sigma_1],\ldots
,[\Sigma_d]$ in $H_2(M,{Z\!\!\!Z})$, corresponding to 2-dimensional
submanifolds in $M$. With the help of the $\mu$-map we get the
corresponding cohomology elements $\mu(\Sigma_i)\in H^2({\cal
  M}_k,{Z\!\!\!Z})$ and can now define the invariant as the pairing
\begin{equation} q_k=\int\limits_{[{\cal M}]}
  \mu(\Sigma_1)\cup\ldots\cup\mu(\Sigma_d) \qquad .  \end{equation}
Because of the non-compactness of the moduli space this definition is
wrong, but as Donaldson showed one can ``repair'' this defect by
evaluating the cup-product $\mu(\Sigma_1)\cup\ldots\cup\mu(\Sigma_d)$
on a regular subset of ${\cal M}_k$. As a result of analysis of the
structure of the Donaldsons invariant of a smooth simply connected
4-manifold, Kronheimer and Mrowka \cite{KrMr:94} proved the existence
of $d$ classes $e_1,\ldots,e_d\in H^2(M,{Z\!\!\!Z})$ and $d$
polynomials $f_1,\ldots, f_d\in\mbox{ \leavevmode\hbox{\kern.3em\vrule
    height 1.2ex depth -.3ex width .2pt\kern-.3em\mbox{\rm Q}}}[z]$
such that the Donaldsons invariant is uniquely defined by these pairs.
Furthermore the set of these classes $\{ e_i\}$ with $i=1,\ldots,k$ is
a diffeomorphism invariant. Suppose a smooth simply connected
4-manifold $M$ with $d$ basic classes is given. A useful realization
of the $\mu$-map is given by the following construction.  We choose a
Dirac operator $/\mkern -10mu \partial_\Sigma$ on $\Sigma$ and the
trivial $SU(2)$ bundle $E$ over $\Sigma$. Next we twist the Dirac
operator with a connection $A$ of $E$ to get $/\mkern -10mu
\partial_{\Sigma,A}$. By ${\cal B}$ we denote the set of all (framed)
connections and by ${\cal E}$ the bundle over $({\cal
  B}/SO(3))\times\Sigma$ induced by $E$. The family index $ind(/\mkern
-10mu \partial_{\Sigma},{\cal E})$ of the Dirac operator can be
considered as a vector bundle over ${\cal B}/SO(3)$. Donaldson defined
a line bundle by
\[ {\cal L}_\Sigma = \bigcup\limits_{A\in{\cal B}/SO(3)}(\Lambda^{max}
\ker/\mkern -10mu \partial_{\Sigma,A})^* \otimes\Lambda^{max}\ker/\mkern -10mu \partial^*_{\Sigma,A} \] i.e.
as determinant bundle of the index bundle $ind(/\mkern -10mu \partial_{\Sigma},{\cal
  E})$.  This line bundle ${\cal L}_\Sigma$ has the following property
\begin{equation} c_1({\cal L}_\Sigma)=\mu([\Sigma]) \qquad .  \end{equation} For more details
see \cite{DonKro:90}. Together with this representation of the
$\mu$-map we can rewrite the Donaldson invariant as
\begin{eqnarray*}
 q_k= \int\limits_{[{\cal M}]} c_1({\cal L}_{\Sigma_1})\cup\cdots\cup
 c_1({\cal L}_{\Sigma_d}) \, .
\end{eqnarray*}
Another construction, which is probably connected with Donaldsons
invariant, was introduced by Seiberg and Witten \cite{SeWi:94,Wit:94}.
To fix the notation we describe the procedure to get the
Seiberg-Witten invariant (for details see \cite{Sal:95}). Because 
of the fact \cite{HiHo:58} that every compact, connected,
orientable,  differentiable four-manifold admits a $Spin^c$-structure 
one choose such a structure together with a section, known as spinor.
Then one write down a system of non-linear PDE describing the coupling
between a self-dual $U(1)$-field and a harmonic spinor. We shall
refer to a $Spin_c$--structure on $M$ by specifying the Spinor Bundle
$S\otimes L$ where $L$ is the square root of a complex line bundle. 
The equations of the gauge theory are given in terms of a pair $(A,\psi)$ of
indeterminates, of which $A$ is a connection on $L$ and
$\psi$ is a smooth section of $S^+\otimes L$.

The equations are
\begin{eqnarray}
D_A\psi &=& 0 \label{eqSW1}\\
(F^+_A)_{ij} &=& \frac{1}{4}<e_ie_j\psi,\psi>e^i\wedge e^j, \label{eqSW2}
\end{eqnarray}
where $D_A$ is the Dirac operator twisted by the connection $A$, and
$F^+_A$ is the self--dual part of the curvature associated to $A$.
Here $\{ e_i \}$ is a local basis of $TX$ that acts on $\psi$ by
Clifford multiplication (see \cite{Sal:95}), $\{ e^i \}$ is the dual
basis of $T^*X$, and $<,>$ is the inner product on the fibres of
$S^+\otimes L$. We denote the moduli space of non-trivial solution of
the Seiberg-Witten equations modulo the gauge group ${\cal
  G}=Map(M,S^1)$ by ${\cal SW}$. Consider the group of all gauge
transformations that fix a base point, ${\cal G}_0\subset {\cal G}$.
Take the moduli space ${\cal SW}_0$ of solutions of Seiberg--Witten
equations, modulo the action of ${\cal G}_0$. This space ${\cal SW}_0$
fibres as a principal $U(1)$ bundle over the moduli space $\cal SW$.
Let ${\cal L}$ denote the line bundle over $\cal SW$ associated to this
principal $U(1)$ bundle via the standard representation.  The
Seiberg--Witten invariant, relative to a choice of $L$ such that the
dimension of $\cal SW$ is positive and even,
\[ 2d=c_1(L)^2-\frac{2\chi+3\sigma}{4}>0, \]
is given by the pairing of the $d$th-power of the Chern class of the
line bundle ${\cal L}$ with the moduli space $\cal SW$,
\[ N_L\equiv \int_{\cal SW} c_1({\cal L})^{d}=\int_{\cal SW} c_d({\cal
  L}^{\oplus d}). \]
If the dimension of $\cal SW$ is odd, the invariant is set to be zero.

There is a conjecture of Witten that the Donaldson invariants and the
Seiberg-Witten invariants are related for manifolds of simple
type. Afterwards Fintushel and Stern \cite{FiSt:95} proved this
conjecture in the case of simple connected elliptic surfaces to be true.
As one expected from the topological classification of four-manifolds,
embedded surfaces and the intersection with another surface
characterize also the differential structure of the manifold. 

\subsection{Changing the Differential Structure}
\label{sec:3.2}
According to the section \ref{sec:2} we introduce two manifolds $M$
and $M^\prime$ with different differential structures and define a map
$\alpha:M\longrightarrow M^\prime$.  Let $TM$ be the tangential bundle
$TM$ of the manifold and $TM^\prime$ for the other case.  In
\cite{KrMr:94} the 4-manifold of simple type was introduced including
the best-known examples like elliptic surfaces (see \cite{FriMor:94}).
Later Fintushel and Stern \cite{FiSt:95,FiSt:95.2} determine the
structure of Donaldson invariants of such 4-manifolds and its relation
to the Seiberg-Witten invariant. Following the discussion in
\cite{KrMr:94} we choose these two differential structures in such a
manner that the number of classes in $H_2(M,{Z\!\!\!Z})$,
characterizing the Donaldson invariant, differs by 1.\footnote{This is
  the simplest case which one can consider as the generator of the
  other cases.} That $M$ has $d$ classes $\Sigma_1,\ldots,\Sigma_d \in
H_2(M,{Z\!\!\!Z})$ and $M^\prime$ has $d+1$ classes with a further
$\Sigma_{d+1}\in H_2(M,{Z\!\!\!Z})$. The interesting point of the map
$\alpha:M\to M^\prime$ is the mapping to the further class
$\Sigma_{d+1}$ of $M^\prime$.  At first we map $d$ classes of $M$ to
$M^\prime$. The further class $\Sigma_{d+1}$ produce a singularity in
the following sense. In \cite{KrMr:93,KrMr:95.2} Kronheimer and Mrowka
investigate the case of an embedded surface in the 4-manifold and its
influence on the Donaldson invariant. As a main result they proved
that except for the case of a sphere the self-interactions
$\Sigma\cdot\Sigma$of the embedded surface $\Sigma$ is given by:
$2g-2\leq \Sigma\cdot\Sigma$ where $g$ is the genus of $\Sigma$.  For
simplicity we neglect self-interactions and choose $g=0$. According to
\cite{KrMr:93} the tangential bundle $T\Sigma$ must be non-trivial. We
choose suitable embeddings $i_1:\Sigma_1\subset M^\prime$ and
$i_{d+1}:\Sigma_{d+1}\subset M^\prime$.  The extension of the bundles
${\cal L}_{\Sigma_1}$ and ${\cal L}_{\Sigma_{d+1}}$ defined in section
\ref{sec:3.1} to the whole manifold $M^\prime$ leads to
\begin{equation}
c_1({\cal L}_{\Sigma_1})\wedge c_1({\cal
  L}_{\Sigma_{d+1}})=c_2((i_1)_\star {\cal
  L}_{\Sigma_1}\oplus (i_{d+1})_\star {\cal L}_{\Sigma_{d+1}})
\end{equation}
and with the definition of the $\mu$-map we get
\begin{equation}
\int\limits_{T_d}\mu([\Sigma_1])\wedge
\mu([\Sigma_{d+1}])=\frac{1}{16}\int\limits_{T_d\times N} p_2({\cal P}^{ad})
\end{equation}
where $N=(i_1)(\Sigma_1)\times (i_{d+1})(\Sigma_{d+1})$ is a
submanifold of $M^\prime$ and $T_d$ is a 4-dimensional submanifold of
the moduli space ${\cal M}^\prime$ of anti-selfdual connections. If
the integral given above is non-trivial then both surfaces $\Sigma_1$
and $\Sigma_{d+1}$ intersect in $M^\prime$. So the existence of the
further class in $M^\prime$ produce further intersections which does
not exists in $M$.  So the interesting points of the map $\alpha$ and
the induced map in the tangential bundle $d\alpha$ are the further
intersection points.  For simplicity we assume that only one further
intersection point between $\Sigma_1$ and $\Sigma_{d+1}$ exists. We
denote this point by $x_0$. The tangential bundle $TM^\prime$ in the
neighborhood $U(x_0)$ of this point splits like
$TM^\prime|_{U(x_0)}=((i_1)_\star T\Sigma_1\oplus (i_{d+1})_\star
T\Sigma_{d+1})|_{U(x_0)}$, because near the intersection point we have
a natural bundle reduction of the rank-4-bundle into two
rank-2-bundles. The connection of $TM^\prime$ in the neighborhood of
this point is the direct sum of the connections in $T\Sigma_1$ and
$T\Sigma_{d+1}$. Because of the non-triviality of the bundles
$T\Sigma_1$ and $T\Sigma_{d+1}$ we obtain also non-trivial connections
in a neighborhood of the intersection point $x_0$. Because of the
relation
\begin{equation}
\oint\limits_{|z|=1} \frac{dz}{z}=2\pi i = \int\limits_K
d\left(\frac{dz}{z}\right)= 2\pi i \int\limits_K
\delta(x)\delta(y)\, dx\wedge dy
\end{equation}
with $z=x+iy\in{\leavevmode\hbox{\kern.3em\vrule height 1.2ex depth
    -.3ex width .2pt\kern-.3em\mbox{\rm C}}}$ and $\partial K=\{
z\in{\leavevmode\hbox{\kern.3em\vrule height 1.2ex depth -.3ex width
    .2pt\kern-.3em\mbox{\rm C}}};\, |z|=1\}$, we can construct a
connection on a nontrivial complex line bundle $T\Sigma_1$ by the map
$\beta:\Sigma_1\times{\leavevmode\hbox{\kern.3em\vrule height 1.2ex
    depth -.3ex width .2pt\kern-.3em\mbox{\rm C}}}\to T\Sigma_1$ with
$\beta(x,1)=a(x)$ and $a(x_0)=0$. This connection is given by the
push-forward of the form $z^{-1}dz$ by the map $\beta$ leading to
$\beta_\star(z^{-1}dz)=a_\star(z^{-1}dz)=a^{-1}da$. In
\cite{HarLaw:93} this case is studied extensively and one obtains for
a connection $\omega$ in $T\Sigma_1$
\begin{eqnarray}
  \omega= \frac{da}{a} = a^{-1}da
\end{eqnarray}
induced from the trivial connection on
$\Sigma_1\times{\leavevmode\hbox{\kern.3em\vrule height 1.2ex depth
    -.3ex width .2pt\kern-.3em\mbox{\rm C}}}$ with the property
$d(a^{-1}da)\not= 0$. The map $d\alpha:TM\longrightarrow TM^\prime$ is
in the neighborhood of the point $y_0$ with $\alpha(y_0)=x_0$ given by
the map of the trivial bundle
$(i_1)_\star(\Sigma_1\times{\leavevmode\hbox{\kern.3em\vrule height
    1.2ex depth -.3ex width .2pt\kern-.3em\mbox{\rm C}}})\times
V\times{\leavevmode\hbox{\kern.3em\vrule height 1.2ex depth -.3ex
    width .2pt\kern-.3em\mbox{\rm C}}}$ with $V$ a two-dimensional
subset of $M$ homeomorphic to a subset of the ${I\mkern-6.0mu R}^2$,
to the nontrivial bundle $((i_1)_\star T\Sigma_1\oplus (i_{d+1})_\star
T\Sigma_{d+1})|_{U(x_0)}$. If we define the splitting of the map
$d\alpha$ in the neighborhood $U(y_0)$ of $y_0$ by the expression
$d\alpha|_{U(y_0)}=(b_1,b_{d+1})$ then the connection induced by the
bundle map is given by $(b_1^{-1}db_1)\oplus(b_{d+1}^{-1}db_{d+1})$.
>From the summable property of the connections near the intersection
point we obtain for the change of the connection the expression
\begin{eqnarray}
\label{change}
  \nabla^\prime = \nabla +
  (b_1^{-1}db_1)\oplus(b_{d+1}^{-1}db_{d+1})\quad .
\end{eqnarray}
This is the final result of the connection change. The additional term
disappears if the manifolds have the same differential structure.  
Next we will interpret the results given above in a physical context.

\section{Interpretation}
In this section we try to give an answer to the following questions:
What is the physical relevance of the differential structure? Which
physical observable will be modified after a change of the
differential structure? In principle the second question can be
answered by the results of the previous section but to learn more
about this topic one has to investigate all invariants of the
differential structure. In most cases these invariants are topological
invariants which no relation to the analytical properties of
manifolds, i.e. the homotopy type or the combinatorical construction
of an invariant. But in most physical models defined by the equation
of motion the geometry of the underlying manifold plays an essential
role in the description and one is faced with the problem to transport
a vector along a curve, to form derivatives of physical observables
and so on. Usually one implicitly fixes the differential structure for
the whole space and all problems are gone.  But in a future version of
quantum gravity all geometrical and topological states of the space
are included in a space of states of the system.

In the general relativity theory (GRT), first of all one fixes the
differential structure of the manifold to be able to write down the
field equation for a description of the gravitational field. This
field equation is a second order non-linear PDE explicitly given on a
chard.  From this fact it follows, that in the GRT only the
infinitesimally generated diffeomorphisms in the chard (or in more
mathematical terms, the diffeomorphisms connected to the identity
component) are used. In physics this means that only the local
gravitational field in a local system (for instance the famous
elevator of Einstein) represented by one chard can be neglected by
acceleration. To follow this effect, Einstein developed a field
equation where the gravitational field is connected with the geometry
of the space. The problem of this theory is the source of the
gravitation given by the energy of the system which is not determined
by the theory itself. So let us consider two different differential
structures leading to different gravitational theories. As in the
previous section we choose two 4-manifolds $M$ and $M^\prime$ with
different differential structures and non-diffeomorphic tangential
bundles $TM$ and $TM^\prime$. Next we consider a connection $\nabla$
of the tangential bundle $TM$ which produces the Riemannian curvature
$R(X,Y)Z=\nabla_X\nabla_Y Z-\nabla_Y\nabla_X Z+\nabla_{[X,Y]}Z$, where
$X,Y,Z\in{\Gamma}(TM)$ are vector fields and ${\Gamma}(TM)$ denotes
the set of all vector fields of $M$. Let $Ric(X,Y)$ be the Ricci
tensor, $R$ the curvature scalar and $g(X,Y)$ the metric defined for
all $X,Y\in{\Gamma}(TM)$. Then Einsteins vacuum field equation is
given by
\begin{equation} 
Ric(X,Y)-\frac{1}{2}g(X,Y)R=0 \quad\mbox{in}\quad M\qquad . \label{einstein1}
\end{equation} 
>From the last section we know that the change of the differential
structure produces a correction (\ref{change}) to the connection
leading to a different curvature. Here we remark that the Lie-algebra
of the Lorentz group $SO(3,1)$ is isomorphic to $sl(2,{\leavevmode\hbox{\kern.3em\vrule
     height 1.2ex depth -.3ex width .2pt\kern-.3em\mbox{\rm C}}})$ which is
the Lie-algebra of the $SL(2,{\leavevmode\hbox{\kern.3em\vrule
     height 1.2ex depth -.3ex width .2pt\kern-.3em\mbox{\rm C}}})$ or spin group. So the connection of
$TM$ can be described by complex $2\times 2$ matrices and the
(real) connection 1-form is denoted by $\omega^j_i$ with $i,j=1,2,3,4$. Together
with the two complex functions $b_1=b_1(z)$ and $b_{d+1}=b_{d+1}(z)$
($z\in{\leavevmode\hbox{\kern.3em\vrule
     height 1.2ex depth -.3ex width .2pt\kern-.3em\mbox{\rm C}}}$), the change of the connection by changing the differential
structure according to the previous section can be written as 
\begin{equation}
\omega^\prime=\omega + \left(
\begin{array}{cc}
\frac{db_1}{b_1} & 0 \\ 0 & \frac{db_{d+1}}{b_{d+1}}
\end{array}\right) \qquad .
\end{equation}
If we denote the connection change by $\Delta\omega$ then one obtains
\begin{equation}
(\omega^\prime)^j_i=\omega^j_i+(\Delta\omega)^j_i=\omega^j_i+\delta^j_i\frac{db^j_i}{b^j_i}
\end{equation}
where $\delta^j_i$ is Kroneckers $\delta$-function and $b^1_1=Re(b_1(z))$,
$b^2_2=Im(b_{1}(z))$, $b^3_3=Re(b_{d+1}(z))$ etc. are
abbreviations. According to the standard 
calculus with $\omega^j_i=\Gamma^j_{ki}dx^k$ and the fact according to
the theory of $U(1)$ connections that
$\Delta\omega\wedge\Delta\omega=0$, we obtain  the correction to the
Ricci tensor $\Delta Ric$ and the scalar curvature $\Delta R$
\begin{eqnarray}
(\Delta Ric)_{ik} &=& \partial_k(\Delta\Gamma)^j_{ji}
-\partial_j(\Delta\Gamma)^j_{ki}\\
\Delta R &=& g^{ik}(\Delta Ric)_{ik}
\end{eqnarray}
with$(\Delta\omega)^j_i=(\Delta\Gamma)^j_{ki}dx^k=(b^j_i)^{-1}db^j_i$.
Here we remark that in the case of the Levi-Civita connection the
metric also change after the change of the differential
structure. From the general relation between the metric tensor and the
Levi-Civita connection one obtains the correction of the metric by
solving the system of differential equations:
\begin{eqnarray*}
  (b^{-1}\partial_i b)^j_j=\frac{1}{2}g^{jm}(\partial_j
  g_{mi}+\partial_i g_{jm} -\partial_m g_{ij}) \quad .
\end{eqnarray*}
Finally we obtain the correction of the vacuum equation
(\ref{einstein1}) by the change of the differential structure
\begin{eqnarray}
(Ric)_{ik}-\frac{1}{2}g_{ik} R &=& \delta^j_i(\partial_j(b^{-1}\partial_k
b)^j_{i}-\partial_k(b^{-1}\partial_j
b)^j_{i}\nonumber\\ &+& \frac{1}{2}g_{ik}(g^{lm}( \partial_j(b^{-1}\partial_l 
b)^j_{m}-\partial_l(b^{-1}\partial_j b)^j_{m})))\label{einstein2}
\end{eqnarray}
But from the relation
\begin{equation}
\oint\limits_{|b|=1} \frac{db}{b}=2i\pi w=2i\pi w\int\limits_K \delta(b)*1
\end{equation}
where $w$ is the winding number or multiplicity of the zero $b(z)=0$,
$\partial K=\{b\in{\leavevmode\hbox{\kern.3em\vrule
     height 1.2ex depth -.3ex width .2pt\kern-.3em\mbox{\rm C}}} ; |b|=1\}$ and $*1$ is the volume form of $K$, we
obtain 
\begin{equation}
\frac{-i}{2 w \pi}d(b^{-1}db)=\delta(b)*1
\end{equation}
a $\delta$-like singularity in the curvature\footnote{Such
  distribution-valued differential forms are known as
  currents in mathematics.}. Together with the basis $\partial$ of the
tangential space with $dx^k(\partial_j)=\delta^k_j$ we obtain for
(\ref{einstein2}):
\begin{eqnarray}
(Ric)_{ik}-\frac{1}{2}g_{ik} R &=& \delta^j_i\{
[(d(b^{-1}db))(\partial_j,\partial_k)]^j_i\nonumber\\ &+& \frac{1}{2}g_{ik}(g^{lm}
[(d(b^{-1}db))(\partial_j,\partial_l)]^j_m \}\quad .
\end{eqnarray}
If we use the abbreviation
$(\delta(b)*1)(\partial_i,\partial_k)=\delta(b)_{ik}$ then we get the
final result
\begin{equation}
(Ric)_{ik}-\frac{1}{2}g_{ik} R=2\pi w
\delta^j_i(\delta(b^j_i)_{jk}+\frac{1}{2}g_{ik}(g^{lm}
\delta(b^j_m)_{jl})
\end{equation}
which means that
\begin{equation}
Ric(X,Y)-\frac{1}{2}g(X,Y)R\not= 0\quad\mbox{in}\quad
M^\prime \qquad .  
\end{equation} 
But the right hand side of this equation represents the
source of the gravitational field given by the energy-momentum tensor.
In our case this term represents the embedding of a surface in a
4-dimensional manifold, because the support of the $\delta$ function
is a two-dimensional manifold. One can interprets such a term as a
embedded (compact) surface in a 4-manifold with the energy given by
the product of the volume of the surface and the winding number of the
map. If one fix one coordinate to be the time coordinate then one can
interpret such a term as a string moving through the 4-dimensional
manifold. So we have the main
result:\\[0.2cm]
{\em The change of the differential structure leads to a singularity
  in the curvature considered in the standard differential
  structure. The change corresponds to a source of a gravitational
  field given by a embedded surface}\\[0.3cm] 
That means if we live in the coordinate system with standard
differential structure and look at a system with an exotic one, we
observe this system with an additional gravitational field
corresponding to the change. In four dimensions this source can be
interpreted as a string. 

\section*{Acknowledgments}
I wish to thank H. Ros\'e for fruitful discussions which clarify the
physical interpretation of differential structures. Special thanks to
G. He{\ss} for many discussions about the work of Harvey and Lawson.
Furthermore I thank Prof. Nietsch and T. Mautsch for many talks about
the mathematical details. Last but not least I want to thank Prof.
Brans for many discussions, his helpful remarks and corrections of
mistakes.

\end{document}